\documentclass[
final            
  ]
  {aipproc}

\usepackage{SIunits}
\usepackage{aas_macros}

\layoutstyle{6x9}

\begin{document}

\title[Microsecond Time Resolution Optical Photometry with H.E.S.S.]
      {Microsecond Time Resolution Optical Photometry using a H.E.S.S. Cherenkov Telescope}


\author{Christoph Deil}{
  address={Max-Planck-Institut f\"ur Kernphysik, P.O. Box 103980, 69092 Heidelberg, Germany}
}

\author{Wilfried Domainko}{
  address={Max-Planck-Institut f\"ur Kernphysik, P.O. Box 103980, 69092 Heidelberg, Germany}
}

\author{German Hermann}{
  address={Max-Planck-Institut f\"ur Kernphysik, P.O. Box 103980, 69092 Heidelberg, Germany}
}

\begin{abstract}
We have constructed an optical photometer with microsecond time resolution, which is currently being operated on one of the H.E.S.S. telescopes.
H.E.S.S. is an array of four Cherenkov telescopes, each with a 107\,\square\meter\ mirror, located in the Khomas highland in Namibia. In its normal mode of operation H.E.S.S. observes Cherenkov light from air showers generated by very high energy gamma-rays in the upper atmosphere.
Our detector consists of seven photomultipliers, one in the center to record the lightcurve from the target and six concentric photomultipliers as a veto system to reject disturbing signals e.g. from meteorites or lightning at the horizon. The data acquisition system has been designed to continuously record the signals with zero deadtime.
The Crab pulsar has been observed to verify the performance of the instrument and the GPS timing system. Compact galactic targets were observed to search for flares on timescales of a few microseconds to $\sim$100\,\milli\second.
The design and sensitivity of the instrument as well as the data analysis method are presented.
\end{abstract}

\maketitle


\section{Introduction}

The motivation for high time resolution photometry is that it provides insight into the astrophysics of the most compact objects in the universe. For example neutron stars and stellar mass black holes with radii $\sim$10\,\kilo\meter\ can show variability on timescales down to tens of microseconds, if the emission comes from the surface (in the case of a neutron star) or from the inner parts of an accretion disk around these objects.

The most prominent example is the Crab pulsar. It has a period of $\sim$33\,\milli\second\ and during its main pulse it is variable in the optical on timescales down to at least 100\,\micro\second\ \cite{1993ApJ...407..276P,2007ApSS.308..595K}.
Besides pulsars, X-ray binaries containing a compact object which is accreting matter from a normal star typically show variability on millisecond timescales since 90\% of the gravitational energy is released in the inner $\sim$100\kilo\meter, mostly in thermal X-rays \citep{2000ARAA..38..717V}. But  evidence for fast optical variability from X-ray binaries has been reported as well (e.g. \cite{1994AA...289..141B}). Compared to X-ray observations, optical variability studies have the advantage that photon collection areas $\sim$100 times larger can be used and that photon energies are $\sim$1000 times smaller. This means that it could be possible to observe variations on timescales where X-ray observations are photon limited.

High time resolution optical observations are usually done with large optical telescopes \cite{2007arXiv0707.2325E,2001ExA....11..157S,2007MNRAS.378..825D}. An alternative is to use Cherenkov telescopes, which have very large reflectors but rather poor angular resolution. Cherenkov telescopes are used to detect very high energy ($\sim$100\,\giga\electronvolt--100\,\tera\electronvolt) gamma-rays by imaging Cherenkov light from air showers produced by these gamma-rays in the upper atmosphere. The regular gamma-ray observations with Cherenkov telescopes cannot take place during moonshine, since scattered moonlight increases the night sky background (NSB) to a level where the faint flashes of Cherenkov light can no longer be detected (note though that MAGIC does perform limited gamma-ray observations during moderate moonshine \citep{2007astro.ph..2475M}). This presents the possibility for optical observations with these telescopes, e.g. regular monitoring programs of variable sources like X-ray binaries.

\section{The H.E.S.S. Cherenkov Telescopes}

\begin{figure}
\includegraphics[width=\textwidth]{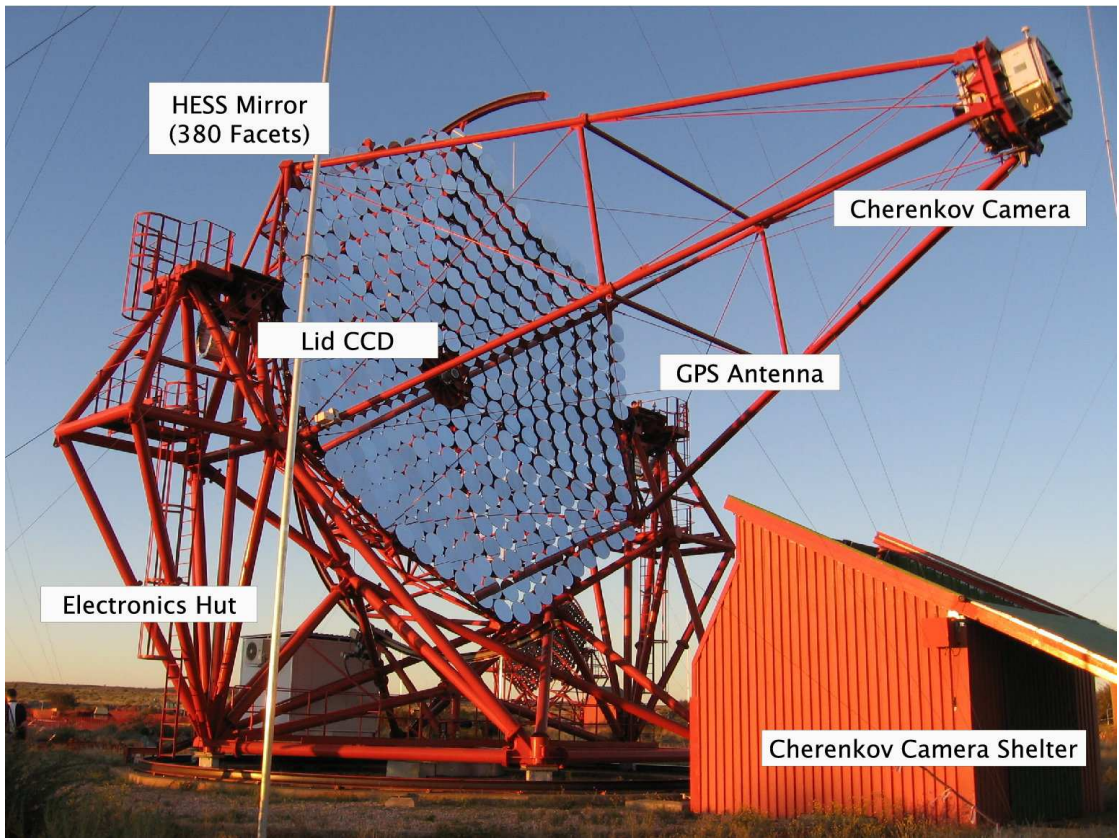}
\caption{The H.E.S.S. telescope CT4 in side view.}
\label{fig:TelescopeSideView}
\end{figure}

\begin{figure}
\includegraphics[width=\textwidth]{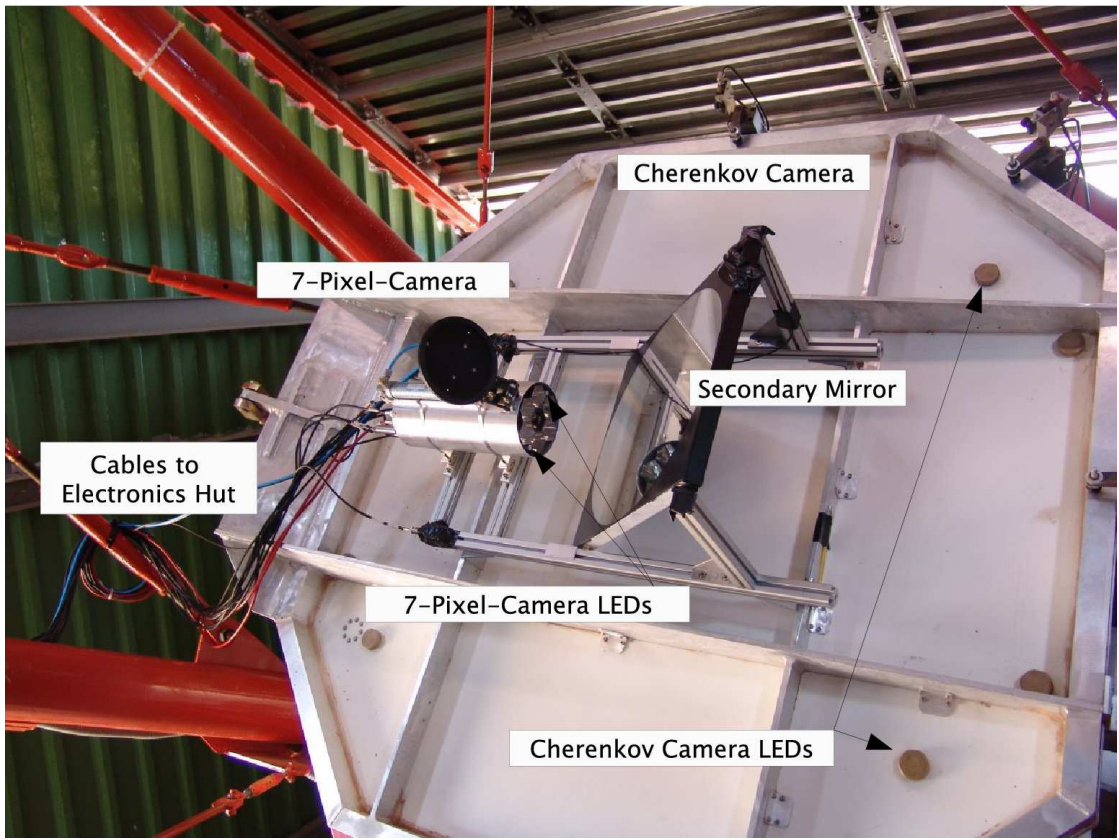}
\caption{The 7-pixel camera mounted on the lid of the Cherenkov camera of CT4.}
\label{fig:7PixCamMounted}
\end{figure}

We have built an optical photometer (similar to the one described in \cite{2006APh....26...22H}) and installed it on the H.E.S.S. telescope CT4 in January and May 2007 (see Figures \ref{fig:TelescopeSideView} and \ref{fig:7PixCamMounted}).

The advantage of using a H.E.S.S. telescope is its large segmented mirror (107\,\square\meter\ area, $\sim$12\meter\ diameter, $\sim$75\% reflectivity) and in comparison to other Cherenkov telescopes its good angular resolution (FWHM$\sim0.042\,\degree=150\,\arcsecond$). An overview of the relevant parameters for optical observations with the four major imaging atmospheric Cherenkov experiments currently in operation is given in Table \ref{tab:IACTs}. The reason why Cherenkov telescopes have poor angular resolution in comparison to optical telescopes (typically FWHM$\sim1$--2\,\arcsecond\ limited by seeing) is that they are designed to image air showers with diameters $\sim$1\,\degree\ and the performance of determining air shower parameters (like length, width and intensity) does not increase significantly with pixel sizes below $\sim0.05\,\degree$ (for a detailed discussion on the H.E.S.S. optical system see \cite{2003APh....20..111B, 2003APh....20..129C}).

\begin{table}
\begin{tabular}{crccrrc}
  \hline
  \tablehead{1}{c}{b}{Name}
  & \tablehead{1}{c}{b}{Location}
  & \tablehead{1}{c}{b}{Start}
  & \tablehead{1}{c}{b}{No. of\\Tel.}
  & \tablehead{1}{r}{b}{Area per\\Tel. [\square\meter]}
  & \tablehead{1}{r}{b}{FWHM[\degree]}
  & \tablehead{1}{c}{b}{Reference} \\
  \hline
  CANGAROO III & 31\degree S 137\degree E& 2003 & 4 & 57  & 0.210 & \citep{2004NewAR..48..323K}\\
  H.E.S.S. & 23\degree S~~~16\degree E & 2003 & 4 & 107  & 0.042 & \citep{2004NewAR..48..331H}\\
  MAGIC & 29\degree N~~18\degree W&2003  & 1  & 236   & 0.082 & \citep{2004NewAR..48..339L} \\
  VERITAS & 32\degree N 111\degree W&2007 & 4 & 110  & 0.060 & \citep{2006astro.ph.11598H}\\
  \hline
\end{tabular}
\caption{Imaging air Cherenkov telescope system parameters relevant for optical observations. Shown is the location, the start of operation of all telescopes, the number of telescopes in the array, the mirror area per telescope and the FWHM of the PSF (of the optical system, not to be confused with the spatial resolution for gamma-ray observations).}
\label{tab:IACTs}
\end{table}

For the focal length of $\sim15\,\meter$ of the H.E.S.S. telescopes the PSF FWHM corresponds to $11\,\milli\meter$ in the focal plane. By applying online pointing corrections (usually done offline for gamma-ray observations) for atmospheric refraction, telescope deformations and the offset of the photometer relative to the center of the Cherenkov camera, a pointing accuracy of less than $40\,\arcsecond$ or $3\,\milli\meter$ was achieved. This allowed us to install an aperture of diameter $22\,\milli\meter$ corresponding to $0.084\,\degree=300\,\arcsecond$ in front of the central pixel, containing $\sim90\%$ of the signal while reducing the night sky background.

\section{The 7-pixel Camera}

\begin{figure}
  \resizebox{.45\textwidth}{!}{\includegraphics{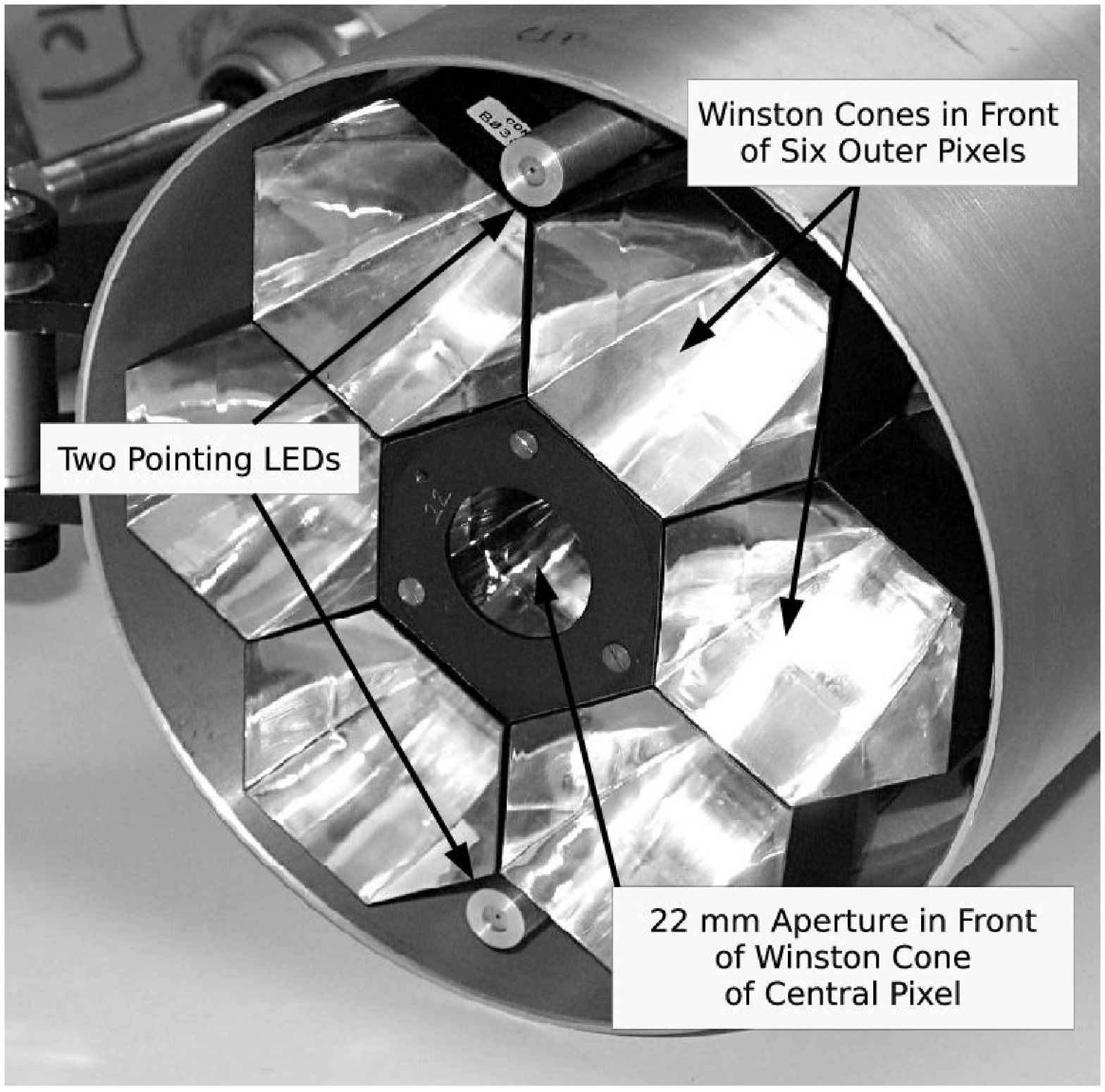}}
  \resizebox{.55\textwidth}{!}{\includegraphics{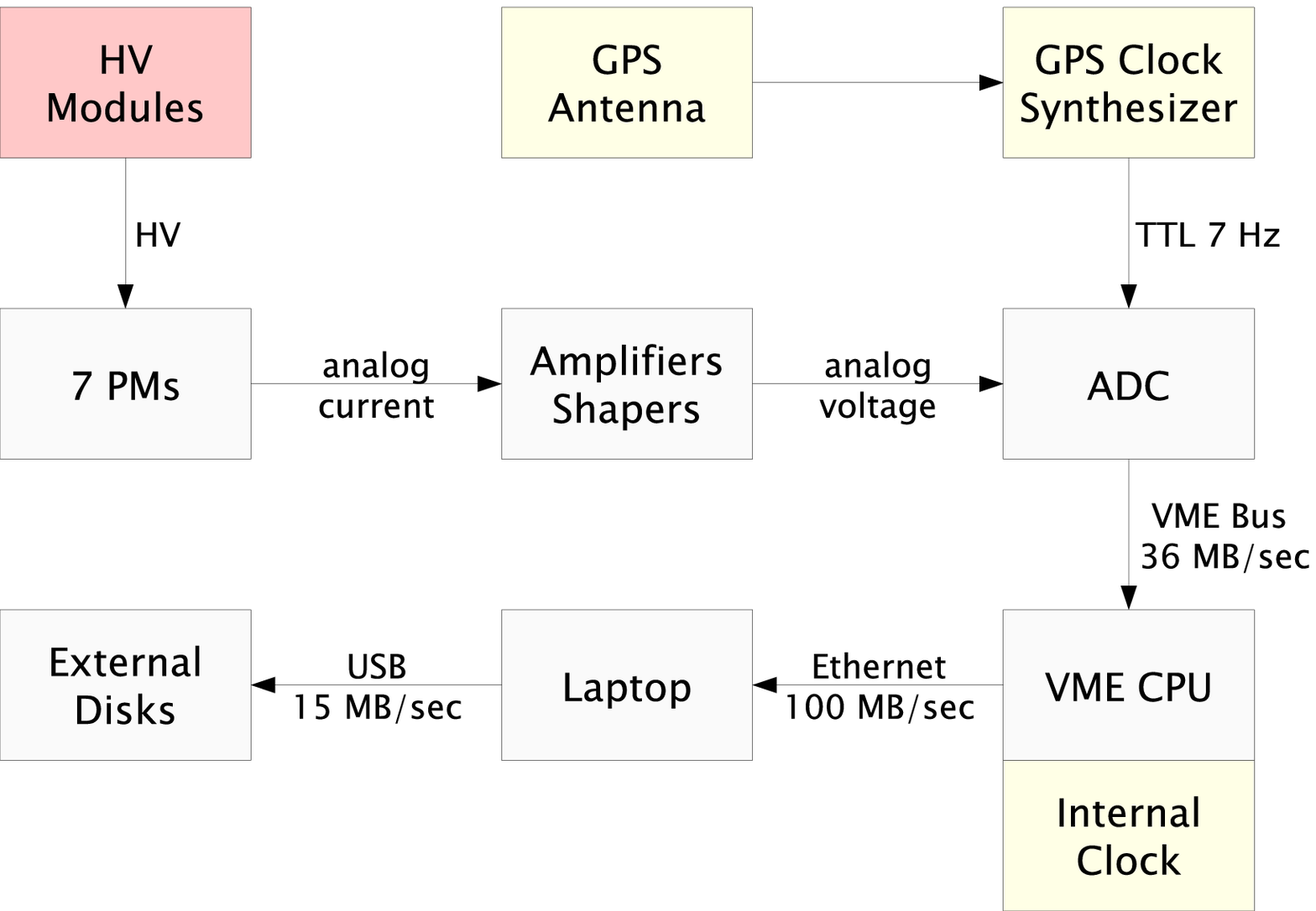}}
\caption{
\textit{Left:} The 7-pixel camera in front view.
\textit{Right:} The 7-pixel camera DAQ system. Also shown are the maximally possible data rates for VME, Ethernet and USB.
}
\label{fig:DAQ}
\end{figure}

We have built a high time resolution photometer consisting of seven photomultipliers (PMs), one to record the light curve from the target and six concentric photomultipliers to reject disturbing signals like meteorites, air showers from cosmic rays, or distant lightning (see Figure \ref{fig:DAQ} (left)).

The H.E.S.S. Cherenkov camera is focused on a height of $10\,\kilo\meter$, since this is the typical height of the air showers it observes. It is constructed such that the focus for stars is on its lid and stellar images, as seen by a CCD camera located in the reflector, can be used to align the H.E.S.S. mirrors. Therefore, to install a second detector, one has to use a plane secondary mirror to put it in the focus (see Figure \ref{fig:7PixCamMounted}). For our setup this meant that the 7-pixel camera had to be unmounted/mounted every night before/after the  Cherenkov camera lid was opened/closed for gamma-ray observations. This could be done in only a few minutes, the mechanical precision of remounting the camera was negligible ($\sim0.3\,\milli\meter$). This was checked every night by measuring the position of two LEDs on the 7-pixel camera relative to the position of 8 LEDs on the Cherenkov camera. To check the pointing accuracy we regularly closed the pneumatic lid of the 7-pixel camera and measured the position of bright stars on this lid in comparison to the 7-pixel camera LEDs.

The data acquisition (DAQ) system used to record the seven lightcurves is shown in Figure~\ref{fig:DAQ} (right). We used Photonis XP2960 photomultipliers (PMs) as light detectors, with hexagonal Winston cones \citep{2003APh....20..129C} as light funnels in front of them to achieve $360\,\degree$ coverage around the central pixel. The PMs were operated at high voltages of 800--1300\volt, depending on the NSB due to moonshine, corresponding to gains $3\times10^4$--$7\times10^5$. The high voltage was chosen such that the mean anode current from background light was always $\sim10$--$20\,\micro\ampere$. These PMs are most sensitive in the wavelength range $\sim270$--$550\,\nano\meter$, with a peak quantum efficiency of $\sim30\%$ at $\sim400\nano\meter$.

We used $\sim$50\,\meter\ cables to lead the PM signals to the electronics hut, where they were amplified by custom-built, three-stage low-noise amplifiers with a 50\,\ohm\ input resistor, gain $\sim1360$ and shaped with a simple $\tau=RC=50\,\ohm\,\cdot\,10\,\nano\farad=0.5\,\micro\second$ low-pass filter. To digitize the signal we used a SIS 3301 Flash ADC which has eight channels, 14\,bit resolution, 5\,\volt\ dynamic range and a sampling speed of 100\,\mega\hertz.

The ADC could in principle record the lightcurves at 10\,\nano\second\ time resolution, but this would generate a data rate of 1.6\,GB\per\second\ (2\,byte\per sample and 8 channels---7 pixels plus an interference monitoring channel connected to a resistor in the 7-pixel camera), which cannot be written to disk. Since we wanted to record the lightcurve continuously, we programmed the ADC to average over 128 consecutive samples, giving an effective sampling speed of 1.28\,\micro\second\ and a data rate of 12\,MB$\per\second=43\,$GB\per h.

In order to get the absolute timing information for every sample, we used a Meinberg LANTIME M300 GPS clock to generate a 7\,\hertz\ TTL signal, where every 7th TTL pulse occured at a full second. The ADC was used in autobankswitch mode, meaning that the recording of the signal was switched between its two internal memory banks every time it was given a TTL signal from the GPS clock. The reason why we used 7\,\hertz\ instead of 1\hertz\ as the bank switch rate is that one memory bank can store 131072 samples and $10^8/128/7\sim111607$ samples easily fit in one bank. While the ADC was recording the signal in one memory bank, the other was simultaneously read out by a VME CPU and written to disk together with a stop timestamp from the internal clock of the VME CPU. Since the precision of this clock was better than 1\,\milli\second\ (it was synchronized to the GPS clock, which also acted as a LAN time server), the absolute time of every sample to the precision of the GPS clock ($\sim1\micro\second$) is known. Putting 7 memory banks together always gives 781236 samples, which shows the excellent stability of the sampling rate of the ADC.

\section{Observations}

The 7-pixel camera was built in November/December 2006 and installed, together with the electronics, on CT4 from Jan 20--27 2007. Due to bad weather conditions, only a few hours of observations were possible. The Crab pulsar was observed to verify the setup and timing system.
Observations were continued from May 2-29 and $\sim$40 hours of observations were performed, mainly on optically  bright low and high mass X-ray binaries. Runs are typically only $\sim$30 minutes long because they were interrupted by pedestal and pointing runs to check the proper working of the system. For future observations, several hours of uninterrupted monitoring are possible.

\begin{figure}
\includegraphics[width=\textwidth]{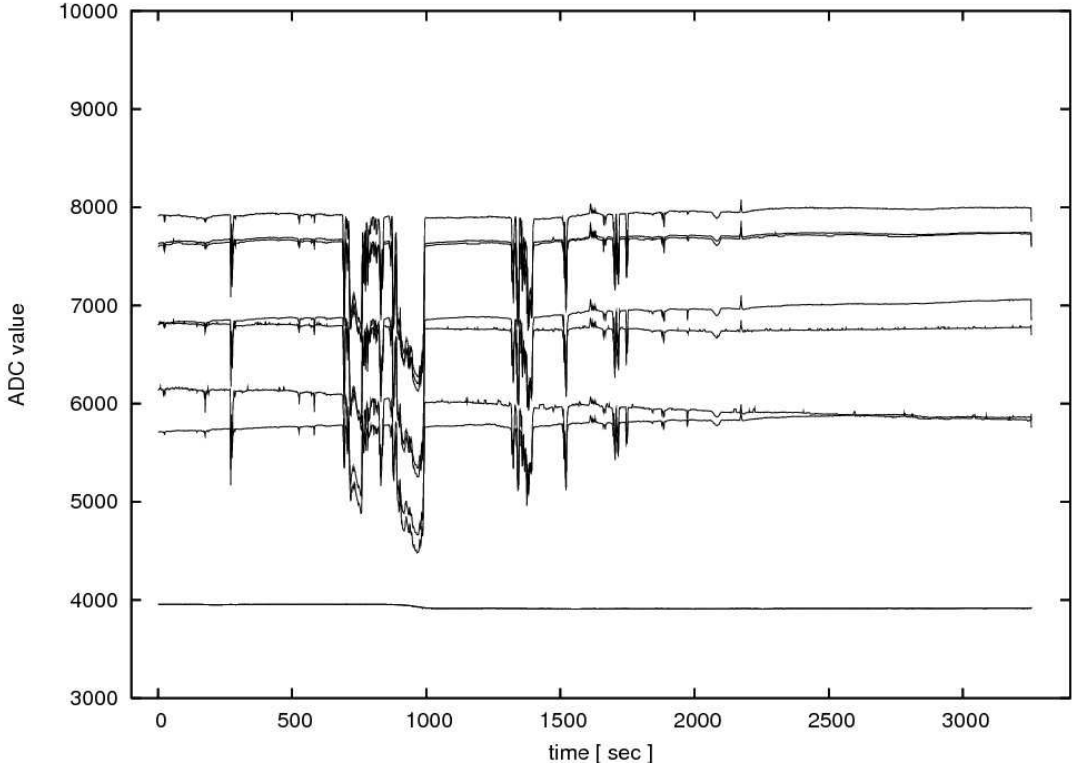}
\caption{Example overview plot for run 0489 on V 4641 Sgr with run duration 54 minutes. Shown is the mean signal of the seven pixels and the interference monitoring channel in 1\second\ intervals. This run was classified as bad because clouds were passing the field of view.}
\label{fig:RunOverview}
\end{figure}

Before searching the data for flares, a quality selection based on the visual inspection of the lightcurves of each run, binned to 1 second resolution, was done (see Figure \ref{fig:RunOverview}). For $\sim$80\% of the May observations, the conditions were good.

\section{Analysis Method}

The data analysis consists of two steps, first the detection of flares in the central pixel and second the veto of coincident flares in the outer pixels. Since the waveform and timescales of variability from the targets are unknown, the flare finding algorithm has to be very general. We chose the peak correlator (PC) algorithm, as described in \cite{1999PhRvD..59h2002A} in the context of the detection of gravitational wave bursts by interferometric detectors. The basic idea is to correlate the signal with a Gaussian $\exp(-t^2/2\tau^2)$ as a generic peak function and to apply a threshold on the correlation. We use the implementation by \cite{BuL}, which is very fast since it calculates the correlations in frequency space, and also takes detector noise (like 50\hertz\ interference from power supplies) into account by estimating and correcting for the power spectral density.

To detect peaks at different time scales, we use $\sim$30 different $\tau$, optimally distributed across the range 1\,\micro\second\,--\,100\,\milli\second. The output of the PC algorithm is a list of so called micro events. Micro events are the largest consecutive intervals where the correlation is above the threshold (typically $\sim$7 standard deviations) and among other parameters have a start and end time $T_s,\ T_e$ and an amplitude $A_{max}$ at time $T_{max}\in[T_s,T_e]$.
Since often several templates with closeby $\tau$ trigger on a flare, this micro event list is clusterized into a macro event list by merging two micro events if they almost overlap, i.e. if $[T_s-T_{tol},T_e+T_{tol}]$ overlaps for a certain tolerance $T_{tol}$, usually chosen to be $T_{tol}\sim5\tau$.

\begin{figure}
\includegraphics[width=\textwidth]{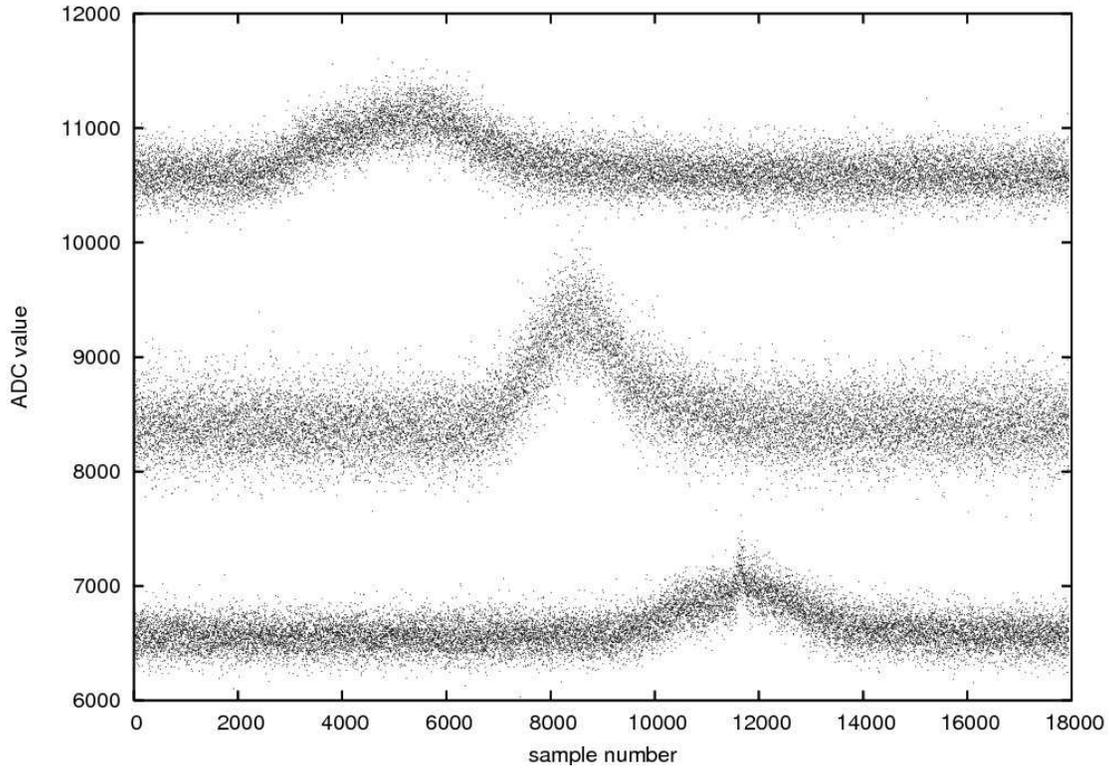}
\caption{Example background event causes by a meteorite. The flare first occured in the outer pixel number 5 (top lightcurve), then in the central pixel (middle lightcurve), and at last  in the outer pixel 1 (bottom lightcurve). The sampling rate of the run was 2.56\,\micro\second, the time delay between the flares is $\sim$3400\,samples or $\sim$8.7\,\milli\second.}
\label{fig:MeteoriteEvent}
\end{figure}

At timescales of 1\,\milli\second--10\,\milli\second, there is quite a large amount of flares in the central pixel caused by meteorites, which are also detected in two or three of the outer pixels with a delay of a few 10\,\milli\second\ (see Figure \ref{fig:MeteoriteEvent}). It is difficult to distinguish these background events from flares actually coming from the astronomical target, since they are time-delayed and do not have the same waveform in the outer pixels (typically longer there since the outer pixels don't have an aperture in front of the Winston cone). Lightning and other terrestial signals are typically simultaneous and have the same waveform, and thus are easy to veto by making a cut on the correlation coefficient of the signal from the central pixel with each of the outer pixels in the time interval $[T_s-T_{tol},T_e+T_{tol}]$.

Because of the meteorites we decided to apply the flare-finding algorithm to all seven pixels and then make a coincidence veto on the macro event lists, where only flares from the central pixel are considered that are at least $\sim$150\,\milli\second\ away from a flare in any of the outer pixels (there seems to be a cutoff in the distribution of meteorite delays at 130\,\milli\second). The analysis of the January and May 2007 data as well as simulations to determine the detection efficiency and false alarm rate are still in progress.

\section{Outlook}

H.E.S.S. is currently building a fifth Cherenkov telescope in the center of the array (planned to be operational in late 2008), which will have a segmented reflector with a mirror area of $\sim600\,\square\meter$---this will be the largest optical reflector in the world---and a PSF FWHM of $\sim0.040\degree$ \citep{2005ICRC....5..171C}. Using the same photometer with that much larger telescope would increase the sensitivity by $\sim2.4$, and another improvement could be made by using high quantum efficiency PMs that better match the optical wavelength region 300--600\,\nano\meter.

It would be desirable to perform simultaneous observations with a second fast photometer, installed either on a normal optical telescope or at another Cherenkov telescope like MAGIC, which would practically eliminate all atmospheric background events.


\begin{theacknowledgments}
We would like to thank Stefan Schmidt and Christian Neureuther for the manufacture of the 7-pixel camera, Thomas Kihm for help with the DAQ design and software, Christopher van Eldik and Oliver Bolz for help with the implementation of the online pointing corrections, Andreas F\"orster, Toni Hanke and Eben Tjingaete with the installation of the instrument and observations in Namibia, and Andr\'e-Claude Clapson for suggesting the PC algorithm and help with the data analysis.
\end{theacknowledgments}

\bibliographystyle{aipproc}   

\bibliography{}

\end{document}